\documentclass[sort&compress,twocolumn,5p]{elsarticle} 

\usepackage{graphicx}

\usepackage{amsmath,amsfonts}

\usepackage{natbib}

\usepackage[pdftex,breaklinks,
            pdftitle={Comments on the Voigt function implementation in the Astropy and SpectraPlot.com packages},
            pdfsubject={Numerical methods and codes; Computational Techniques},
            pdfkeywords={Complex error function; Complex probability function; Plasma dispersion function; Fadde(y)eva function},
            pdfauthor={F. Schreier}]{hyperref}

\batchmode

\newcommand{\D}{\mathrm{d}}
\newcommand{\E}{\mathrm{e}}
\newcommand{\I}{\mathrm{i}}
\newcommand{\qufig}[1] {Fig.~\ref{#1}}

\journal{J.\ Quant.\ Spectr.\ \& Rad.\ Transfer, received 28 Feb 2018, accepted 23 March 2018; doi: 10.1016/j.jqsrt.2018.03.019}

\begin{document}
\begin{frontmatter}

\title{\textsf{Notes} \\[1ex]
       Comments on the Voigt function implementation in the Astropy and SpectraPlot.com packages}
\author{Franz Schreier}

\address{DLR --- Deutsches Zentrum f\"ur Luft- und Raumfahrt, \\ 
           Institut f\"ur Methodik der Fernerkundung,  \\
           82234 Oberpfaffenhofen, Germany}
\cortext[ca]{Corresponding author}
\ead{franz.schreier@dlr.de}

\begin{abstract}
The Voigt profile is important for spectroscopy, astrophysics, and many other fields of physics, but is notoriously difficult to compute.
\citeauthor{McLean94} [J.\ Electron Spectrosc.\ \& Relat.\ Phenom., 1994] have proposed an approximation using a sum of Lorentzians.
Our assessment indicates that this algorithm has significant errors for small arguments.
After a brief survey of the requirements for spectroscopy we give a short list of both efficient and accurate codes and recommend implementations based on rational approximations.
\end{abstract}



\end{frontmatter}


\section{Introduction}
\label{sec:introduction}

The Voigt profile, the convolution of a Lorentzian and Gaussian profile \citep{NIST-handbook,DLMF}, is ubiquitous in many branches of physics including spectroscopy and astrophysics \citep[e.g.][]{Heng17b,Bernath16}.
Because there is no closed-form solution for this integral, numerous algorithms have been developed using series or asymptotic expansions, continued fractions, Gauss-Hermite quadrature,
etc. (see \citet{Armstrong67} for an old but still interesting review and section \ref{sec:discussion} for a more extensive survey).

Several approximations of the Voigt function have been proposed using empirical combinations of the Lorentz or Gauss functions
\citep{McLean94,Matveev72,Wertheim74,Whiting68,Kielkopf73,Martin81,Puerta81,Bruce00,Liu01}.
Although numerous sophisticated algorithms have been developed, this type of approximation appears to be still quite popular.
Recently, the \citet{McLean94} approximation has been implemented in the ``\texttt{SpectraPlot.com}'' \citep{Goldenstein17} and ``\texttt{astropy}'' \citep{astropy13etal} packages.
Unfortunately, this approach has serious accuracy problems esp.\ for small arguments and is suboptimal w.r.t.\ efficiency.

The objective of this comment is to demonstrate these shortcomings and to present accurate and efficient alternatives.
We first review some basic facts, present the performance of the \citeauthor{McLean94} approximation in Section \ref{sec:results}, and discuss alternatives in Section \ref{sec:discussion}.

\section{Theory}
\label{sec:theory}

The Voigt profile \citep{Armstrong67} is defined by
\begin{align}
 g_\text{V}(\nu-\hat\nu,\gamma_\text{L},\gamma_\text{G})
       &=~ \int_{-\infty}^\infty \D\nu' \; g_\text{L}(\nu-\nu',\gamma_\text{L}) \,\times\,
                                           g_\text{G}(\nu'-\hat\nu,\gamma_\text{G})  \\
       &=~ {\sqrt{\ln 2 / \pi} \over \gamma_\text{G}} ~K(x,y) \\
 \intertext{with the closely related Voigt function}
 K(x,y) ~&=~ {y \over \pi} ~ \int_{-\infty}^\infty {\E ^{-t^2} \over (x-t)^2 + y^2} ~ \D t ~,
\end{align}
where $\gamma_\text{L}$ and $\gamma_\text{G}$ are the half widths at half maximum (HWHM) of the Lorentzian $g_\text{L}$ and Gaussian $g_\text{G}$, respectively,
and $\hat\nu$ is the center wavenumber or frequency.
The arguments of the Voigt function are defined as
\begin{equation}\label{defxy}
 x ~=~ \sqrt{\ln 2}~ {\nu - \hat\nu \over \gamma_\text{G}} 
\qquad\hbox{ and }\qquad
 y ~=~ \sqrt{\ln 2}~ {\gamma_\text{L} \over \gamma_\text{G}} ~.
\end{equation}
Note that the Lorentz, Gauss, and Voigt profiles are normalized to one, whereas for the Voigt function $\int K \,\D x = \sqrt\pi$.

The Voigt function comprises the real part of the complex error function (a.k.a.\ complex probability function, Fadde(y)eva function, or plasma dispersion function)
\begin{equation}\label{wDef}
 w(z) ~\equiv~ K(x,y) \,+\, \I L(x,y) ~=~ {\I \over \pi} ~ \int_{-\infty}^\infty ~ {e^{-t^2} \over z-t} ~ \D t ~.
\end{equation}
with  $z = x + \I y$.
The complex error function satisfies the differential equation
\begin{equation} \label{diffEqu}
w'(z) ~=~ -2z\cdot w(z) \:+\: {2\I \over \sqrt\pi} ~.
\end{equation}

\citet{McLean94} suggested an approximation based on \citet{Puerta81} using a sum of four Lorentzians,
\begin{equation} \label{mcLean}
 K(x,y) ~=~ \sum_{l=1}^4 {c_l(y-a_l) + d_l(x-b_l) \over (y-a_l)^2 + (x-b_l)^2}
\end{equation}
where $a_1$, \dots, $d_4$ are real constants. 
Ignoring the sign, there are only eight distinct parameters to ensure the symmetry $K(-x,y)=K(x,y)$.
For an accuracy test \citeauthor{McLean94} compared the approximation with the convolution integral evaluated numerically for five $y$ between $0.1$ and $10$ and found a maximum absolute error $3.65 \cdot 10^{-4}$.


\section{Results}
\label{sec:results}

For an assessment of the code performance it is important to know the range of $x$ and $y$ to be expected.
For infrared line-by-line atmospheric radiative transfer calculations \citet{Wells99} encountered $|x| < 33\,000$ and $10^{-4}< y < 125$.
\citet{Lynas-Gray93} discussed requirements for astrophysical spectroscopy and expects $y\le 1$ for stellar absorption.
For spectroscopy of terrestrial planetary atmospheres $y$ can be as small as $10^{-8}$ \citep{Schreier11v}.

Furthermore, the accuracy required for function evaluation depends on the accuracy of the input data.
In view of the quality of spectroscopic line parameter databases such as HITRAN \citep{Gordon17etal} or GEISA \citep{JacquinetHusson16etal}, four significant digits have been considered as adequate \citep{Schreier11v}.

\begin{figure}
 \centering\includegraphics[width=0.5\textwidth]{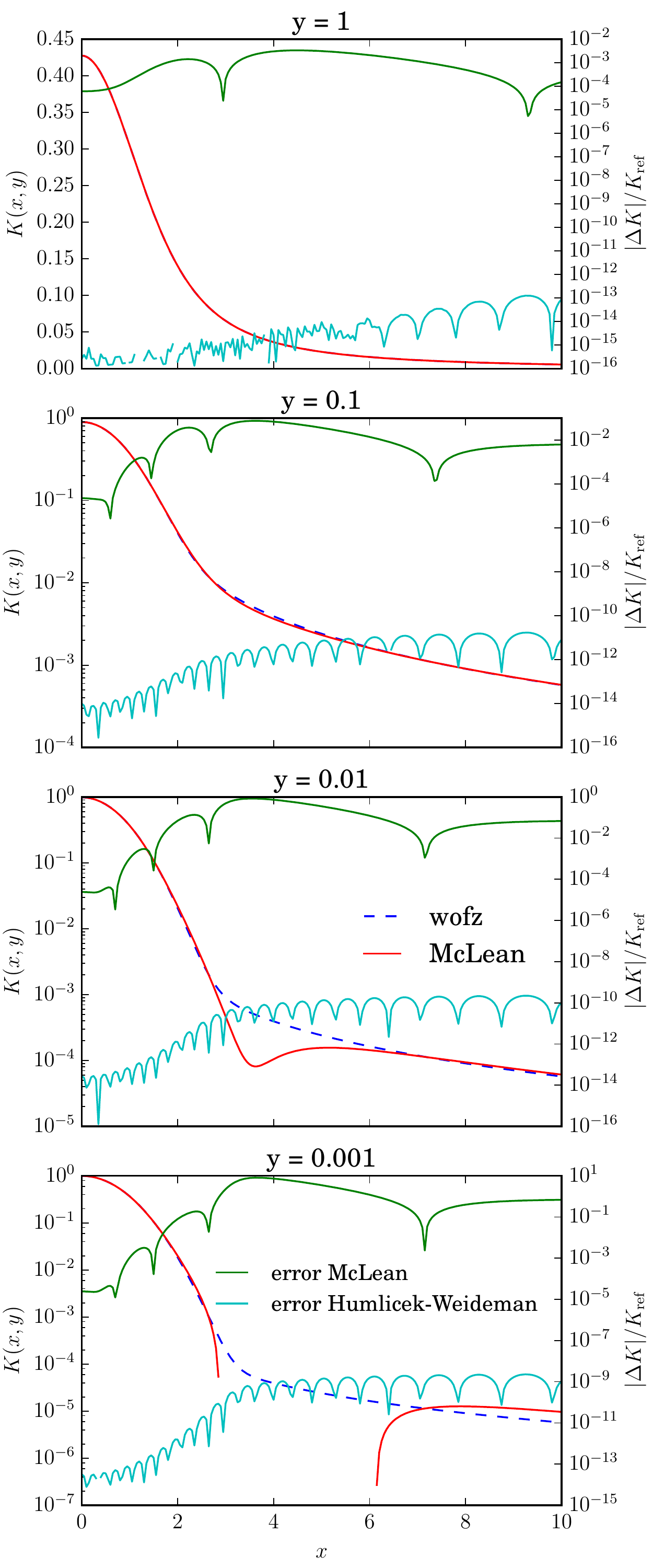}
 \caption{Comparison of the McLean approximation (red) with the \texttt{wofz} reference (blue dashed).
          Left $y$-axis: Voigt function values for $y=1$ (top) to $y=0.001$ (bottom);
          Right $y$-axis: relative error of the McLean (green) and \citeauthor{Humlicek82}--Weideman (cyan) algorithms.}
 \label{fctError}
\end{figure}

For the evaluation of the accuracy of the \citet{McLean94} approach we used SciPy's \texttt{wofz} function
(\texttt{scipy.special.wofz}, a combination of the \citet{Poppe90a,Poppe90} and \citet{Zaghloul11} codes with a stated accuracy of at least 13 significant digits).
In \qufig{fctError} we compare the McLean function values with the real part of the complex error function returned by \texttt{wofz}
and also show the relative error $|K_\text{mcLean} - K_\text{wofz}| / K_\text{wofz}$.
For $y$=1 the two profiles cannot be distinguished, and the relative error is less than $10^{-2}$.
Discrepancies become apparent for $y=0.1$, and for smaller $y$ the failure of the McLean implementation around $x \approx 4$ is obvious.
For $y=0.001$ the relative error reaches almost ten and $K_\text{mcLean}$ becomes negative for $3 \le x \le 6$.
Note that for small $y$ the Voigt function resembles a Gaussian in the center and turns into a Lorentzian in the wings.
Obviously, the \citeauthor{McLean94} approximation especially fails in the transition region, and \qufig{fctError} also indicates problems in the asymptotic region.

\begin{figure}
 \centering\includegraphics[width=0.5\textwidth]{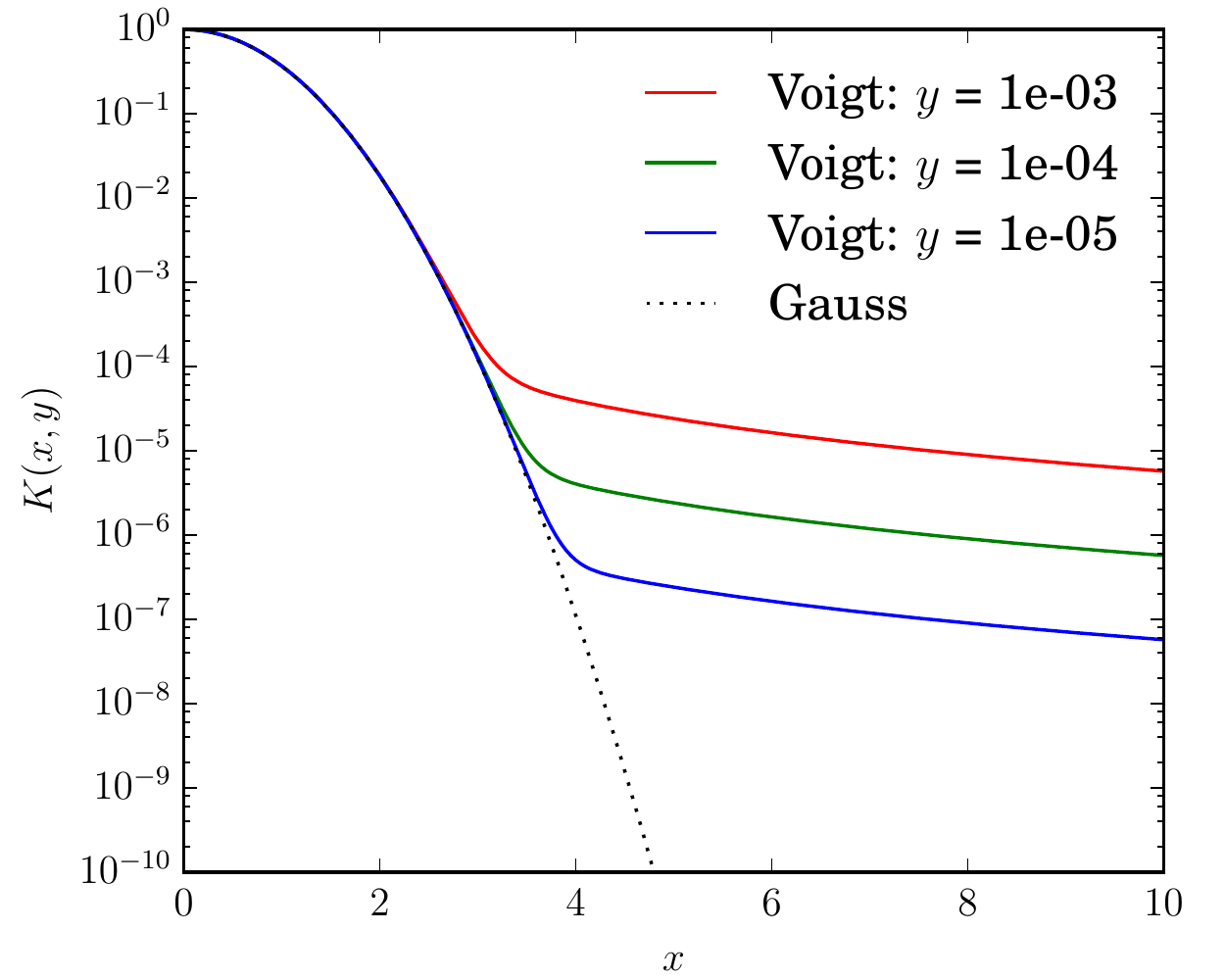}
 \caption{Comparison of the Voigt and Gauss function for small $y$.}
 \label{voigt_gauss}
\end{figure}

The \citet{McLean94} approximation as implemented in the \texttt{SpectraPlot.com} package \citep{Goldenstein17} is used for moderate $0.001 \le y \le 1000$ only, and the Gaussian and Lorentzian is used for smaller or larger $y$, respectively.
However, because asymptotically the Voigt function behaves as $y/(\sqrt{\pi} x^2)$, a Gauss approximation leads to a significant underestimate in the line wings, see \qufig{voigt_gauss}.

\texttt{astropy.Voigt1D} offers the partial derivatives based on analytical differentiation of the approximation \eqref{mcLean}.
However, given the complex error function $w$, the derivatives can be readily evaluated according to \eqref{diffEqu} as
\begin{equation} \label{dKdXY}
 \begin{aligned}
  \partial K(x,y) / \partial x &=~ -2 (xK - yL) \\
  \partial K(x,y) / \partial y &=~ 2 (xL + yK) - {2 \over \sqrt\pi} ~.
 \end{aligned}
\end{equation}
This is especially advantageous for efficiency, because evaluation of the differentiated approximation \eqref{mcLean} requires numerous time consuming divisions.


\section{Discussion}
\label{sec:discussion}

As indicated, dozens of codes have been developed for evaluation of the Voigt and/or complex error function.
For benchmarks highly accurate implementations are available, e.g., \citet[][available at \url{https://dl.acm.org/citation.cfm?doid=77626.77630}]{Poppe90,Poppe90a} providing fourteen significant
digits or \citet[][available at \url{https://dl.acm.org/citation.cfm?id=2049679}]{Zaghloul11} with up to 20 digits.
The accuracy of the \citet{Weideman94} rational approximation can be adjusted by specifying the degree $N$ of the polynomial,
with $N=32$ the relative error with respect to \texttt{wofz} is smaller than $10^{-6}$.
\citet[][see also \url{http://arblib.org/}]{Molin11}  and \citet[][available as online supplement]{Boyer14} have developed algorithms for arbitrary precision.

For spectroscopic applications the Voigt function has to be evaluated for thousands or even millions or billions of lines and for thousands or millions of frequency grid points \citep{Rothman10,Tennyson16etal}.
Naturally, the accuracy of input data is limited, and efficiency becomes more important.

One of the most widely used algorithms for efficient and moderately accurate computation is by \citet{Humlicek82} providing four significant digits (further variants and optimizations by,
e.g., \citet{Kuntz97,Ruyten04,Imai10}.)
Because this \texttt{w4} code uses four different rational approximations for small and/or large $x$ and $y$, it can be difficult to implement efficiently in, e.g., Numeric Python.
Accordingly we have recommended a combination of \citeauthor{Humlicek82}'s asymptotic rational approximation for large $|x|+y$ with Weideman's approximation otherwise \citep{Schreier11v}.
Besides additions and multiplications only one division per function value is needed for both approximations.
This algorithm can be implemented in Numeric Python with less than 10 statements and with $N=32$ has a relative error less than $8\cdot 10^{-5}$ for $0\le x\le 20$ and $10^{-6}\le y
\le 10^2$ (also see \qufig{fctError}.) 

The \texttt{cpf12} function by \citet{Humlicek79} employs two different rational approximations and has a maximum relative error less than $5 \cdot 10^{-6}$.
Whereas the region I approximation can be optimized easily \citep{Schreier11v}, an optimization of the region II approximation is not straightforward.
\citet{Wells99} combined the two \citet{Humlicek79,Humlicek82} approximations and achieves a relative accuracy of $10^{-5}$.
\citet{Lynas-Gray93} provided an optimized implementation of the \citet{Lether90} algorithm for vector machines accurate to 10 decimal digits in the $10^{-7}<y<10^3$ domain
(available at \url{http://cpc.cs.qub.ac.uk/summaries/ACLT_v1_0.html}).
\citet{Shippony93} combined series, rational approximations and Gauss-Hermite integrations and state a maximum relative error of less than $1 \cdot 10^{-8}$ over the complex plane
(note, however, that the plots of the relative error were provided for $0 \le y \le 6$ on a linear axis).
The \citet{Zaghloul17} code allows the accuracy to be adjusted and falls back to \citet{Humlicek82} in case of low accuracy demands (available at \url{https://dl.acm.org/citation.cfm?id=3119904}).

The increasing quality of molecular spectra has indicated deficiencies of the Voigt profile due to line mixing, speed-dependent effects, or Dicke narrowing.
\citet{Tennyson14} provided a thorough discussion and recommended the ``Hartmann-Tran'' profile.
The speed-dependent Voigt profile can be calculated as the difference of two complex error functions (with one of the imaginary parts as small as $10^{-8}$).
For the Hartmann-Tran profile a further term involving the difference of scaled complex error functions has to be considered.
Accordingly, the number of function evaluations and the accuracy requirements increase with more sophisticated line profiles \citep{Schreier17}.


\section{Conclusions}
\label{sec:conclusions}

The implementation of the Voigt function $K(x,y)$ in the popular \texttt{astropy} and \texttt{SpectraPlot.com} packages has been assessed.
The code, based on the \citet{McLean94} approximation, works reasonably well for large $y$, but significantly fails for $y \ll 1$ with relative errors larger than one.
(The Python script defining the Voigt functions and generating the figures can be downloaded from our department's web site at \url{https://atmos.eoc.dlr.de/tools/lbl4IR/mcLean.py}.)

Most modern algorithms compute the complex error function: the real part gives the Voigt function, and derivatives are provided ``on-the-fly''.
A brief survey of some implementations has been given;
in particular a combined \citeauthor{Humlicek82}-\citeauthor{Weideman94} algorithm allowing efficient \emph{and} accurate function evaluations over a large range of the parameter space is recommended
\citep[][available at \url{https://atmos.eoc.dlr.de/tools/lbl4IR/}]{Schreier11v}.


\paragraph{Acknowledgements}
Financial support by the DFG project SCHR 1125/3-1 is greatly appreciated.

\bibliography{JOURNALS,math,physics,radiation,molec,planets,comp,voigt}



\end{document}